\documentclass{article}
\usepackage{spconf}

\usepackage{mathtools,amsmath,amssymb}
\usepackage{graphicx}
\usepackage{booktabs}
\usepackage[tight-spacing=true]{siunitx}
\usepackage{flushend}
\usepackage{subcaption}
\usepackage{csquotes}
\usepackage{xfrac}
\usepackage[hidelinks]{hyperref}
\usepackage[capitalize]{cleveref}
\usepackage[nolist,nohyperlinks]{acronym}

\usepackage{microtype}
\usepackage[subtle]{savetrees}
\newcommand{\tens}[1]{\mathbf{#1}}
\newcommand{\sequence}[1]{\left[#1\right]}

\newcommand{\iter}{\ensuremath{j}}
\newcommand{\iterstotal}{\ensuremath{J}}
\newcommand{\scanidx}{\ensuremath{k}}
\newcommand{\scanidxtotal}{\ensuremath{K}}
\newcommand{\scanidxnew}{\ensuremath{F}}

\newcommand{\object}{\ensuremath{O}}
\newcommand{\probe}{\ensuremath{P}}
\newcommand{\scanpos}{\ensuremath{\mathbf{r}}}
\newcommand{\exitwave}{\Psi}
\newcommand{\exitwaves}{\tens{\exitwave}}

\newcommand{\ampls}{\tens{A}}
\newcommand{\intensities}{\tens{I}}
\newcommand{\est}[1]{\hat{#1}}

\newcommand{\ii}{\mathrm{i}}
\newcommand{\conj}[1]{#1^{*}}

\newcommand{\ftsymb}[0]{\mathcal{F}}
\newcommand{\ft}[1]{\ftsymb\{#1\}}
\newcommand{\iftsymb}[0]{\mathcal{F}^{-1}}
\newcommand{\ift}[1]{\iftsymb\{#1\}}

\newcommand{\segment}{\ensuremath{S}}
\newcommand{\objupdate}{\ensuremath{S^{\dagger}}}
\newcommand{\frozen}{\ensuremath{\text{frozen}}}
\newcommand{\fluid}{\ensuremath{\text{fluid}}}
\newcommand{\epsC}{\varepsilon_\constrC}

\newcommand{\buffersize}{\ensuremath{B}}

\newcommand{\partially}{\text{part}}

\newcommand{\constrA}{\mathcal{A}}
\newcommand{\constrC}{\mathcal{C}}

\newcommand{\proj}[1]{\mathcal{P}_{#1}}
\newcommand{\projA}{\proj{\constrA}}
\newcommand{\projC}{\proj{\constrC}}
\newcommand{\projCpartial}{\projC^{\partially}}

\newcommand{\partialsum}{\ensuremath{s}}
\newcommand{\Owk}{\partialsum^{\probe\exitwave}}
\newcommand{\Wsk}{\partialsum^{|\probe|^2}}

\newcommand{\eqcomma}{\,,} %
\newcommand{\eqperiod}{\,.} %
\newcommand{\minitimes}{{\mkern-2mu\times\mkern-2mu}}

\usepackage[backend=biber, bibencoding=utf8,
            style=ieee,
            maxbibnames=99,
            doi=false,
            url=false,
            isbn=false]{biblatex}
\AtEveryBibitem{%
  \ifentrytype{inproceedings}{%
    \clearfield{pages}%
    \clearlist{publisher}%
    \clearlist{organization}%
    \clearlist{location}}{}%
  \ifentrytype{misc}{%
    \clearfield{eprintclass}{} }}

\title{Live Iterative Ptychography with projection-based algorithms}
\name{Simon Welker$^{1,2}$, Tal Peer$^{1}$, Henry N. Chapman$^{2,3,4}$, Timo Gerkmann$^{1}$\thanks{This work was funded by DASHH (Data Science in Hamburg - HELMHOLTZ Graduate School for the Structure of Matter) - Grant- No. HIDSS-0002, and by 'CUI: Advanced Imaging of Matter' of the Deutsche Forschungsgemeinschaft (DFG) - EXC 2056 - project ID 390715994.}}
\address{\small
    $^{1}$ Signal Processing (SP), Universität Hamburg, Hamburg, Germany\\\small
    $^{2}$ Center for Free-Electron Laser Science CFEL, Deutsches Elektronen-Synchrotron DESY, Notkestr. 85, 22607 Hamburg, Germany\\\small
    $^{3}$ The Hamburg Centre for Ultrafast Imaging, Hamburg, Germany\\\small
    $^4$ Department of Physics, Universität Hamburg, Hamburg, Germany 
}
\begin{document}
\begin{acronym}
\acro{stft}[STFT]{Short-Time Fourier Transform}
\acro{dft}[DFT]{Discrete Fourier Transform}
\acro{rtisi}[RTISI]{Real-Time Iterative Spectrogram Inversion}
\acro{ap}[AP]{Alternating Projections}
\acro{er}[ER]{Error Reduction}
\acro{raar}[RAAR]{Relaxed Averaged Alternating Reflections}
\acro{dm}[DM]{Difference Map}
\acro{gla}[GLA]{Griffin-Lim Algorithm}
\acro{fgla}[FGLA]{Fast Griffin-Lim Algorithm}
\acro{ler}[LER]{Live Error Reduction}
\acro{ldm}[LDM]{Live Difference Map}
\acro{pr}[PR]{Phase Retrieval}
\acro{psnr}[PSNR]{Peak Signal-to-Noise Ratio}
\end{acronym}

\ninept
\maketitle
\begin{abstract}
In this work, we demonstrate that the ptychographic phase problem can be solved in a live fashion during scanning, while data is still being collected. We propose a generally applicable modification of the widespread projection-based algorithms such as Error Reduction (ER) and Difference Map (DM). This novel variant of ptychographic phase retrieval enables immediate visual feedback during experiments, reconstruction of arbitrary-sized objects with a fixed amount of computational resources, and adaptive scanning. By building upon the \emph{Real-Time Iterative Spectrogram Inversion} (RTISI) family of algorithms from the audio processing literature, we show that live variants of projection-based methods such as DM can be derived naturally and may even achieve higher-quality reconstructions than their classic non-live counterparts with comparable effective computational load.
\end{abstract}
\begin{keywords}
Ptychography, phase retrieval, live feedback, X-ray microscopy, diffractive imaging
\end{keywords}

\section{Introduction}
\label{sec:intro}
Ptychography has become established as a popular computational imaging method in the recent past \cite{rodenburgPtychography2019}. Ptychography has several key strengths: It does not necessarily require lenses nor extensive prior knowledge about the imaged specimen, it can retrieve the full complex-valued transmission function of the imaged object, it can reliably recover the illuminating beam (\emph{probe}) along with the imaged object, and it can be applied for imaging problems with X-rays, visible light, and electron beams \cite{rodenburgPtychography2019}. The key idea behind ptychography is to scan an illuminating beam over small overlapping parts of the object and to collect a set of diffraction patterns, one for each scan position. To reconstruct the object, the associated complex phases of the intensities of the diffraction patterns must be computationally recovered, which is the traditionally ill-posed and ill-conditioned \acf{pr} problem. The informational redundancy of the ptychographic scanning process creates enough informational redundancy to make solving this difficult problem comparatively easy and reliable.

In a typical ptychography setup, the scan is first completed and stored on disk, and subsequently processed as a whole. If any experimental parameters were suboptimal, one would then need to adjust them and repeat the whole scan, thus potentially wasting precious beamtime or damaging the sample through excessive exposure. In this work, we observe that the sequential nature of the ptychographic scanning process in principle naturally allows for \emph{live} reconstruction, i.e., a reconstruction process that continually provides a view of the partially scanned object while the scan is still ongoing. This allows for immediate visual feedback, enabling readjustment or termination of the scan when problems become visible.

There are also other works exploring live ptychographic reconstruction, which are however largely limited to specific experimental variants such as Single-Sideband Ptychography \cite{strauchLiveProcessingMomentumResolved2021} or Wigner Distribution Deconvolution \cite{bangunWignerDistributionDeconvolution2023}. In a related recent preprint \cite{weberLiveIterativePtychography2023}, the authors derive a live-capable scheme by making use of the ePIE algorithm \cite{maidenImprovedPtychographicalPhase2009}. Compared to \cite{weberLiveIterativePtychography2023}, here we contribute a different scheme with fixed computational complexity, employ a simple but effective phase initialization strategy, and show that one can also derive live-capable variants of all projection-based methods.

Our contributions in this work are as follows: \textbf{1)} We show how one can extend techniques from the audio processing literature \cite{beauregardEfficientAlgorithmRealtime2005,zhuRealTimeSignalEstimation2007} to derive efficient, fixed-budget live variants of the widespread projection-based algorithms \ac{dm} \cite{elserPhaseRetrievalIterated2003,elserSearchingIteratedMaps2007}, \ac{er} \cite{fienupReconstructionObjectModulus1978} and related methods \cite{welkerDeepIterativePhase2022,lukeRelaxedAveragedAlternating2004,perraudinFastGriffinLimAlgorithm2013}; \textbf{2)} we devise strategies to achieve stable reconstructions when the probe must be retrieved as well; \textbf{3)} we perform extensive evaluations on several varied simulated test objects and probes, and show that our method can not only perform live reconstruction but frequently achieves better reconstructions than its classic non-live counterparts.

\vspace{-2mm}
\section{Methods}
\label{sec:methods}
\subsection{Ptychography}
Ptychography is a family of computational imaging techniques that exploit informational redundancy to solve the problem of \ac{pr} without requiring strong prior knowledge such as the imaged object having compact support or being real-valued.
Ptychography is in particular known for having the ability to solve for the illuminating probe and the illuminated object simultaneously.
We focus here on the case of \emph{far-field ptychography} since it is a two-dimensional analog of the \ac{stft} \cite{welkerDeepIterativePhase2022} in audio processing, for which the \ac{rtisi} algorithms \cite{beauregardEfficientAlgorithmRealtime2005,zhuRealTimeSignalEstimation2007} were developed.
Our proposed methods can in principle also be adapted to variants such as near-field ptychography and Fourier ptychography, which are for instance described in \cite{rodenburgPtychography2019}.

In far-field ptychography, an illuminating probe $\probe$ is scanned over the object $\object$ and interacts with it at a discrete set of positions, which are chosen so that there is a degree of overlap between neighboring illuminated regions. A pixel detector is placed in the far-field downstream of the object and, for each scan position $\scanpos_\scanidx$, records the Fourier intensities $\intensities_\scanidx = |\ft{\exitwave_\scanidx}|^2$ of the \emph{exit wave} $\exitwave_\scanidx(\scanpos) = \probe(\scanpos - \scanpos_\scanidx)\object(\scanpos)$, where $\ftsymb$ is the Fourier transform.
This generates a dataset consisting of a sequence of diffraction patterns $\sequence{\intensities_\scanidx}_{\scanidx=1}^{\scanidxtotal}$ and an associated scan position sequence $\sequence{\scanpos_\scanidx}_{\scanidx=1}^{\scanidxtotal}$. The highly redundant nature of the dataset can then be exploited to solve for both the complex object transmission function $\object$ and the probe function $\probe$.
Typically, an entire such dataset is first collected and subsequently processed and reconstructed as a whole. A popular family of algorithms to perform this reconstruction process are iterative projection-based methods, which are described in the following.

\subsection{Iterative projection-based algorithms}
Iterative projection-based algorithms for diffractive imaging seek to satisfy two or more \emph{constraints} that represent prior knowledge about the target object or about measured detector intensities recorded when illuminating the object (or parts of it). Each constraint can be enforced via an associated projection operator $\proj{}$, which projects an object estimate onto the set of all objects fulfilling the respective constraint. In diffractive imaging problems including ptychography, there are typically only two constraints: a constraint $\constrA$ of the calculated diffraction intensities, and a constraint $\mathcal O$ about the object. In ptychography specifically, $\mathcal O$ takes the form of a \emph{consistency} constraint $\constrC$, which requires all exit waves of the scanned object to be consistent in the overlapping regions. We will thus use the two sets $\constrA, \constrC$ from here on. Their associated projection operators can be written as:
\begin{align}
    \label{eq:pA}
    \projA(\sequence{\exitwave_\scanidx}_{\scanidx=1}^\scanidxtotal) &= \sequence{\ift{\ampls_\scanidx \exp(\ii\ \arg(\ft{\exitwave_\scanidx}))}}_{\scanidx=1}^\scanidxtotal\eqcomma\\
    \label{eq:pC}
    \projC(\sequence{\exitwave_\scanidx}_{\scanidx=1}^{\scanidxtotal}) &= \segment(\objupdate(\sequence{\exitwave_\scanidx}_{\scanidx=1}^{\scanidxtotal}))\eqcomma\\
    \label{eq:object-update}
    \text{with}\ \objupdate(\sequence{\exitwave_\scanidx}_{\scanidx=1}^{\scanidxtotal}) &= \frac{
        \sum_{\scanidx=1}^\scanidxtotal \conj{\est\probe}(\scanpos - \scanpos_\scanidx) \exitwave_\scanidx(\scanpos)
    }{
        \sum_{\scanidx=1}^\scanidxtotal | \est\probe(\scanpos - \scanpos_\scanidx) |^2%
    }\eqcomma\\
    \label{eq:segment}
    S(\est{\object}) &= [\est\probe(\scanpos - \scanpos_\scanidx) \est{\object}(\scanpos)]_{\scanidx=1}^{\scanidxtotal}\eqperiod
\end{align}
where $\conj{}$ is the complex conjugate, $\ampls_\scanidx := \sqrt{\intensities_\scanidx}$ are the measured detector amplitudes, $\est\object$ and $\est\probe$ denote estimates of the object and probe. In practice, we clip the denominator to have a minimum value of $\epsC = 10^{-12}$ everywhere to avoid numerical instabilities where $\probe$ is close to zero.
For $\projC$, the segmentation function $\segment$ cuts up an object estimate $\est{\object}$ into segments and multiplies each with the probe, and its Moore-Penrose pseudoinverse $\objupdate$ creates a consistent object estimate by overlap-adding all segments at their respective scan positions and dividing out the summed probe intensities. $\objupdate$ is also referred to as the \emph{object update} in ptychography literature \cite{rodenburgPtychography2019}. One may write $\segment$ and $\objupdate$ as linear transformations in matrix form by treating the object and probe as vectors. Still, we use a sequence-based notation here since it will make our later derivations using subsequences of exit waves clearer.

The simplest projection-based method is called \acf{er} and is also known as \ac{ap} since it applies the two projections in an alternating fashion:
\begin{equation}\label{eq:alg-er}
    \exitwaves^{\iter+1} \gets \projA(\projC(\exitwaves^{\iter}))
\end{equation}
where $\exitwaves = \left[\exitwave_\scanidx\right]_{\scanidx=1}^{\scanidxtotal}$ denotes the collection of all exit waves. \ac{er} has been shown to guarantee local convergence \cite{fienupReconstructionObjectModulus1978}, but empirically often fails to produce a high-quality reconstruction when used in isolation as it gets stuck in local minima. More advanced methods have thus been developed, such as \acf{dm} \cite{elserPhaseRetrievalIterated2003,elserSearchingIteratedMaps2007}:
\begin{gather}\label{eq:alg-dm}
    \exitwaves^{\iter+1} \gets \exitwaves^\iter + \beta [
        \projA(f_\constrC(\exitwaves^\iter)) - \projC(f_\constrA(\exitwaves^\iter))
    ]\eqcomma
    \\
    f_\constrC(\exitwaves) = \projC(\exitwaves) + (\projC(\exitwaves) - \exitwaves) / \beta\eqcomma
    \\
    f_\constrA(\exitwaves) = \projA(\exitwaves) - (\projA(\exitwaves) - \exitwaves) / \beta\eqperiod
\end{gather}
where $\beta \in \mathbb{R}$ is a tunable parameter. While theoretical guarantees on its convergence are unavailable, \ac{dm} has proven in practice to reconstruct images of significantly higher quality than \ac{er} and today serves as the \emph{de facto} standard workhorse in established ptychography software such as PyNX \cite{favre-nicolinPyNXHighperformanceComputing2020} and PtyPy \cite{endersComputationalFrameworkPtychographic2016}. Other closely related projection-based algorithms such as \ac{raar} \cite{lukeRelaxedAveragedAlternating2004}, \ac{fgla} \cite{perraudinFastGriffinLimAlgorithm2013}, and deep learning methods such as \cite{welkerDeepIterativePhase2022} can also be adapted to our live setting \cite{peerFlexibleOnlineFramework2023}, but we found that using our live variant of \ac{dm} with $\beta=1$ (LDM1) already achieves good reconstructions. We thus leave the exploration of other algorithms and parameters to future work.

\subsection{Real-Time Iterative Spectrogram Inversion}
As discussed in \cite{welkerDeepIterativePhase2022,peerGriffinLimImprovedIterative2022a,kobayashiAcousticApplicationPhase2022,melnykPhaseRetrievalShortTime2023}, there is a direct analogy between \emph{\ac{stft} \acl{pr}} \cite{griffinSignalEstimationModified1984,sturmelSignalReconstructionSTFT2011} in signal processing, and ptychographic \ac{pr}. The \ac{stft} is a widespread time-frequency signal processing technique. It decomposes a signal into overlapping segments, multiplies each segment by a tapered window function to avoid aliasing, and takes the \ac{dft} of each windowed segment.
The key observation here is that a far-field ptychography setup is a physical realization of an \ac{stft}, but with a complex-valued two-dimensional object and probe rather than a real-valued one-dimensional time signal, and typically with less knowledge of---and regularity to---the window (probe) and segment offsets (scan positions) than for a usual \ac{stft}. 
This analogy motivates us to devise ptychographic adaptations of a real-time-capable method for \ac{stft} \ac{pr}, \emph{\acf{rtisi}} \cite{beauregardEfficientAlgorithmRealtime2005,zhuRealTimeSignalEstimation2007}. \ac{rtisi} was developed as a real-time variant of the classic \ac{gla} method \cite{griffinSignalEstimationModified1984}, which is equivalent to \ac{er} and requires an entire spectrogram. %
Derivations of methods similar to ours in the context of \ac{stft} \ac{pr} can be found in parallel work \cite{peerFlexibleOnlineFramework2023}.

\subsection{Live projection-based ptychography}\label{sec:live-ptycho}
\begin{figure}
    \centering
    \includegraphics[width=\columnwidth]{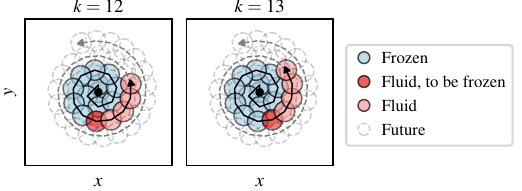}
    \caption{The idea of our method: Moving through the scan positions $\scanidx$ sequentially, we iterate over a fixed-size local buffer of size $B=5$ of \emph{fluid} exit waves while keeping them consistent with all already-reconstructed \emph{frozen} exit waves. After applying an adapted projection-based algorithm on the buffer $\iterstotal$ times, we move one scan position forward and freeze the oldest exit wave in the buffer. Future scan positions are not considered until they are loaded into the buffer.}
    \label{fig:method-idea}
\end{figure}

The derivations in the \ac{rtisi} papers \cite{beauregardEfficientAlgorithmRealtime2005,zhuRealTimeSignalEstimation2007} implicitly assume the use of \ac{gla} as the underlying iterative method, but since \ac{gla} is equivalent to \ac{er}, this would go against common wisdom in ptychography as \ac{er} is unlikely to produce satisfactory results on its own.
We show in the following how, with our notion of a \emph{partial consistency projection}, the idea of \ac{rtisi} can be extended to derive live variants of \ac{dm} and \ac{er}. Other related live algorithms are discussed in our recent preprint for \ac{stft} phase retrieval \cite{peerFlexibleOnlineFramework2023}.
Analogously to \ac{rtisi}, we consider the following scenario, illustrated in \cref{fig:method-idea}: We have reconstructed the first $\scanidxnew$ exit waves to a sufficient degree, so we do not need to update them anymore. We denote these exit waves as $\exitwaves_\frozen = \sequence{\exitwave_\scanidx}_{\scanidx=1}^{\scanidxnew}$. We allocate a \emph{reconstruction buffer} of $\buffersize \geq 1$ exit waves that are still to be reconstructed, $\exitwaves_\fluid = \sequence{\exitwave_\scanidx}_{\scanidx=\scanidxnew+1}^{\scanidxnew+\buffersize}$. Now, for a fixed number of $\iterstotal$ iterations, we update $\exitwaves_\fluid$ using a projection-based algorithm of our choice. We then freeze the oldest exit wave $\exitwave_{\scanidxnew+1}$, use it to update the partial object and probe estimates, and shift the buffer forward by one scan position.

This algorithm is capable of performing \emph{live} ptychographic reconstruction as it only relies on the first $(\scanidxnew+\buffersize)$ exit waves at any given time, and can continually provide an estimate of the partially scanned object. When a new diffraction pattern is recorded at a new scan position, it is queued to eventually become part of the reconstruction buffer. The reconstruction process then continues until the scan is finished. As we use a fixed number of iterations and a fixed buffer size, the computational complexity to reconstruct each single exit wave is independent of the total number of exit waves. This enables live reconstruction of arbitrary-sized objects with a fixed computational budget and also opens the door for adaptive scanning \cite{edeAdaptivePartialScanning2021} that optimizes the next scan positions dependent on the partial object.

However, we must devise a way to keep $\exitwaves_\fluid$ in agreement with $\exitwaves_\frozen$ for all object pixels where there is overlap between the two sets, without updating $\exitwaves_\frozen$. Applying the consistency projection $\projC$ in \eqref{eq:pC} only to $\exitwaves_\fluid$ is not sufficient, as this would not use the information contained in $\exitwaves_\frozen$. Therefore, we propose the notion of a \emph{partial consistency projection} $\projCpartial$, which we derive by truncating the sums in the numerator and denominator of the object update \eqref{eq:object-update} at the $(\scanidxnew+\buffersize)$-th term and splitting each sum into two partial sums:
\begin{equation}\label{eq:partial-object-update}
    \objupdate_\partially\left(\exitwaves_\fluid \left| \exitwaves_\frozen \right.\right) =
    \frac{
        \overbrace{
            \textstyle\sum_{\scanidx=1}^{\scanidxnew} \conj{\est\probe}_\scanidx \exitwave_\scanidx
        }^{\Owk_\frozen}
        +
        \textstyle\sum_{\ell=\scanidxnew+1}^{\scanidxnew+\buffersize} \conj{\est\probe}_{\ell} \exitwave_{\ell}
    }{
            \underbrace{
                \textstyle\sum_{\scanidx=1}^{\scanidxnew} | \est\probe_\scanidx |^2
            }_{\Wsk_\frozen} +
            \textstyle\sum_{\ell=\scanidxnew+1}^{\scanidxnew+\buffersize} | \est\probe_\ell |^2
    }
\end{equation}
where, for brevity, we write $\est\probe_\scanidx(\scanpos) := \est\probe(\scanpos - \scanpos_\scanidx)$ and omit the positions $\scanpos$. The partial sums $\Owk_\frozen$ and $\Wsk_\frozen$ will not change, so only the new $\buffersize$ additional terms must be computed during each algorithm iteration. Similarly, there is no need to calculate new estimates for the past $\scanidxnew$ exit waves, so we define the segmentation operator $\segment_\partially$ to calculate updated versions only of a subsequence of exit waves:
\begin{equation}\label{eq:partial-segment}
    \segment_{\partially}(\est{\object}, \sequence{\scanpos_\scanidx}_{\scanidx=a}^{b}) = [\est\probe(\scanpos - \scanpos_\scanidx) \est{\object}(\scanpos)]_{\scanidx=a}^{b}\eqperiod
\end{equation}
We can now define the \emph{partial consistency projection} as:
\begin{equation}\label{eq:partial-pC}
    \projCpartial(\exitwaves_\fluid | \exitwaves_\frozen)
    :=
    \segment_\partially(\objupdate_\partially(\exitwaves_\fluid | \exitwaves_\frozen), \sequence{\scanpos_\scanidx}_{\scanidx=\scanidxnew+1}^{\scanidxnew+\buffersize})\eqcomma
\end{equation}
and propose that a live-capable variant of any projection-based algorithm such as \ac{dm} \eqref{eq:alg-dm} and \ac{er} \eqref{eq:alg-er} can be derived by replacing each occurrence of the full consistency projection $\projC$ with its partial variant $\projCpartial$. We refer to these variants as \ac{ler} and \ac{ldm} in the following.

\subsection{Informed phase initialization}\label{sec:phase-initialization}
The \ac{rtisi} authors \cite{beauregardEfficientAlgorithmRealtime2005,zhuRealTimeSignalEstimation2007} observe that when a new frame is added to the buffer, we could either follow a typical uninformed strategy and initialize it by, e.g., setting all its Fourier phases to zero, or we could make use of the already-reconstructed parts of the signal that the new frame overlaps with. We can apply the same idea to exit waves in our buffer, by determining an initial phase $\Phi_\scanidx^0$:
\begin{equation}\label{eq:phase-init-phi0}
    \Phi_{\scanidx}^0 = \arg(\ft{\segment_\partially(\est\object, \scanpos_\scanidx)})\eqcomma
\end{equation}
and setting the corresponding initial exit wave estimate to be
\begin{equation}\label{eq:phase-init-psi0}
    \Psi_{\scanidx}^0 = \ift{\ampls_\scanidx \exp(\ii \Phi_{\scanidx}^0)}\eqperiod
\end{equation}
We will see in our experiments that this phase initialization strategy has a strong advantage over the naïve one for reconstruction quality. Note that we still initialize the first $\buffersize$ frames in an uninformed way before the first iteration due to no past information being available.

\subsection{Live probe update}\label{sec:live-probe-update}
A typical ptychographic problem requires solving for the probe and the object simultaneously. It has been shown \cite{thibaultProbeRetrievalPtychographic2009} that a coupled pair of equations, also known as the object update and probe update \cite{rodenburgPtychography2019}, describes a solution to this problem. While the object update is given in \eqref{eq:object-update}, the probe update formula is intuitively obtained by treating the object as the probe and vice versa, resulting in:
\begin{equation}\label{eq:probe-update}
    \est{\probe}(\sequence{\exitwave_\scanidx}_{\scanidx=1}^{\scanidxtotal}) = \frac{
        \sum_{\scanidx=1}^\scanidxtotal \conj{\est\object}(\scanpos-\scanpos_\scanidx) \exitwave_\scanidx(\scanpos)
    }{
        \sum_{\scanidx=1}^\scanidxtotal | \est\object(\scanpos - \scanpos_\scanidx) |^2
    }
\end{equation}
As for the partial object update \eqref{eq:partial-object-update}, a \emph{partial probe update} should make use of the fluid exit waves while not changing the influence of the frozen exit waves on the probe estimate. For this, analogously to \eqref{eq:partial-object-update}, we split the sums in the numerator and the denominator into two partial sums each, a frozen and a fluid partial sum, where the frozen partial sum again does not need to be recalculated for each iteration.

\section{Experimental evaluation}
\label{sec:experiments}

\subsection{Experimental setting}
To explore the performance of the algorithm over a range of imaging conditions we construct a simulated complex-valued test set of 30 distinct objects and scans, with object arrays of size $512\minitimes{}512$ each with a different probe in arrays of size $64\minitimes{}64$. As the basis for our objects, we use 40 randomly chosen RGB images from the DIV2K \cite{agustssonNTIRE2017Challenge2017,timofteNTIRE2017Challenge2017} validation set, resized and cropped to the central square region. We use the following three procedures and construct 10 objects with each: \textbf{1)} convert RGB to HSV (hue/saturation/value), use hue as phase and the value as the amplitude; \textbf{2)} convert RGB to grayscale, use the values (times $2\pi$) as phase and 1 as the amplitude (phase-only objects); \textbf{3)} randomly choose an image pair, convert both to grayscale, use one as phase and the other as amplitude. To further increase the data variety, we transform the amplitudes as $\frac{|\object|+a}{1+a}$ with uniformly random $a \in [0,1]$, and the phases by multiplying with uniformly random $b \in [0.3, 0.99]$. As the probes, we simulate aberrated focused beams: We sample random coefficients of a Zernike polynomial with maximum degree 4, each from a Gaussian distribution with a standard deviation of 0.2 except the Piston coefficient which we set to 0. We evaluate the polynomial on a circular aperture of random fractional radius between $\sfrac{1}{7}$ and $\sfrac{1}{11}$ within a $64\minitimes64$ array and take the array's \ac{dft} to get the probe. For each scan, we use 1500 positions along an Archimedes spiral with polar coordinates $(r, \theta)$ fulfilling $r=10\theta$ (in pixels). This results in 30 scans which have an average overlap ratio \cite{huangOptimizationOverlapUniformness2014} of $\sigma \approx 55.1\%$.

We use our proposed live variant of \ac{dm} with $\beta=1$, referred to as \emph{LDM1}, for all live reconstructions unless stated otherwise. When writing \emph{DM1}, we refer to the established non-live \acl{dm} algorithm with $\beta=1$. To compare reconstruction quality, we use the normalized mean-squared error metric $E_0(\object,\est\object)$ \cite{maidenImprovedPtychographicalPhase2009} evaluated on the central $300\minitimes300$ region of each object. Before calculating error metrics, we zero any phase ramps in $\est\object$ as in PyNX \cite{favre-nicolinPyNXHighperformanceComputing2020}.

\subsection{Optimal parameter estimation}\label{sec:exp:optimal-params}
\begin{figure}[t]
    \centering
    \includegraphics[width=\columnwidth]{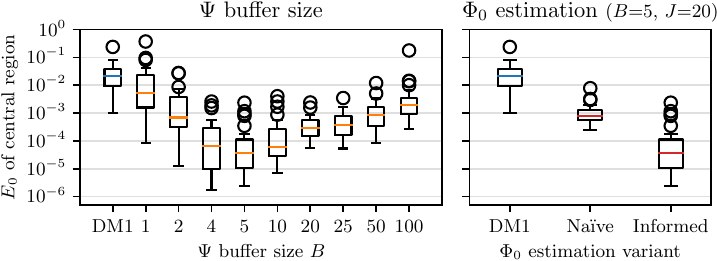}
    \caption{Error metric $E_0$ for LDM1 on all test objects in different scenarios, assuming a known probe. \enquote{DM1} is classic DM with $\beta=1$ and 100 iterations. \textbf{Left:} Changing the buffer size $\buffersize$ while keeping the number of effective iterations per position fixed to $\iterstotal \buffersize = 100$. \textbf{Right}: Phase initialization strategies for the newest exit wave, using optimal $\iterstotal$ and $\buffersize$ from the left. \emph{Naïve}: zero-phase initialization; \emph{Informed}: initialization making use of the past object estimate.}
    \label{fig:lvsj-gtprobe}
\end{figure}

For the experiments in this subsection, we consider the probe to be perfectly known. In our method described in \cref{sec:live-ptycho}, we perform $\iterstotal$ iterations on the current buffer, and then shift the buffer by one exit wave. Thus, each exit wave (except the first $\buffersize$ exit waves) is part of the reconstruction buffer $\buffersize$ times and receives $\iterstotal\buffersize$ effective iterations. We evaluate the optimal setting for $\buffersize$ under the constraint that $\iterstotal\buffersize=100$. In the left part of \cref{fig:lvsj-gtprobe}, we see that, while the effect of different $\buffersize$ can be substantial and $\buffersize=5$, $\iterstotal=20$ performs best, our LDM1 method has a significantly lower reconstruction error than the non-live classic DM1 for almost all settings of $\buffersize$.
For the right-hand part of \cref{fig:lvsj-gtprobe}, we then use this determined optimal setting to test the effect of the phase initialization introduced  in \cref{sec:phase-initialization}. We compare uninformed zero-phase initialization (\emph{Naïve}) with the strategy proposed in \cref{sec:phase-initialization} (\emph{Informed}), see \eqref{eq:phase-init-psi0}. We find that without informed phase initialization, our method already performs better than non-live DM1, but with the proposed phase initialization it has a substantial advantage. This shows that the informed strategy is highly effective.

\vspace{-1.5mm}
\subsection{Probe retrieval}\label{sec:exp:probe-retrieval}
\begin{figure}[t]
    \centering
    \includegraphics[width=\columnwidth]{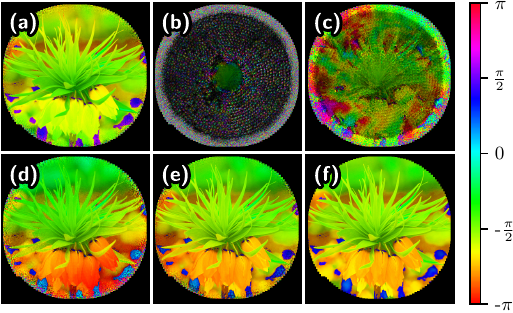}
    \caption{Comparison of reconstructions of an example object, shown in \textbf{(f)}, using different methods while performing probe retrieval. Pixel brightness encodes amplitude, color encodes phase. \textbf{(a)} Classic DM1 with 100 iterations; \textbf{(b)-(e)} Live DM1 (LDM1) with $\iterstotal=10, \buffersize=10$, each variant building upon the previous one: \textbf{(b)} Direct implementation, \textbf{(c)} with quantile clipping, \textbf{(d)} with pre-reconstruction of central region; \textbf{(e)} splitting the 10 iterations into 8x LDM1 and 2x LER. This final live variant achieves a better reconstruction than DM1.}
    \label{fig:alg-progression}
\end{figure}

\begin{table}
\centering
\begin{tabular}{lccc}
\toprule
 & DM1 & LDM1 & LDM1+LER \\
\midrule
$E_0 \downarrow$ & 0.97 $\pm$ 0.18 & 0.34 $\pm$ 0.42 & \textbf{0.11 $\pm$ 0.24} \\
$E_0^\text{c} \downarrow$ & 0.23 $\pm$ 0.29 & 0.12 $\pm$ 0.15 & \textbf{0.11 $\pm$ 0.13} \\
PSNR $\uparrow$ & 4.60 $\pm$ 6.22 & 22.98 $\pm$ 17.56 & \textbf{37.24 $\pm$ 17.42} \\
PSNR$^\text{c} \uparrow$ & 20.48 $\pm$ 11.58 & 27.34 $\pm$ 8.15\phantom{0} & \textbf{27.75 $\pm$ 7.51\phantom{0}} \\
\bottomrule
\end{tabular}
\caption{$E_0$ and the \ac{psnr} of the amplitudes, evaluated on all reconstructed objects from each method when also performing probe retrieval. We report empirical means and standard deviations. LDM1 refers to the final variant of live DM, and LDM1+LER is a hybrid variant combining LDM1 and LER, see \cref{sec:exp:probe-retrieval}. Best values in bold. Superscript $^\text{c}$ denotes the metric being evaluated on the central $300\minitimes300$ region.}
\label{tab:metrics-final}
\end{table}

With the derivations we have made in \cref{sec:live-ptycho} and \cref{sec:live-probe-update}, we now consider the typical ptychographic setting where the probe is unknown and must be reconstructed along with the object. For this, we set the first initial probe estimate to an array filled with complex random Gaussian noise, scaled to the same $\ell_2$ norm as the brightest diffraction pattern in the first buffer, see \cite[Sec.~17.9.3]{rodenburgPtychography2019}.

As direct application of the live variant of the probe update may lead to highly unstable behavior, see \cref{fig:alg-progression}(b), we found it helpful to apply the following modifications: \textbf{1)} We keep the $\ell_2$ norm of the probe estimate constant throughout the reconstruction to eliminate the amplitude ambiguity between object and probe \cite[Sec.~17.4.2]{rodenburgPtychography2019}. \textbf{2)} After finding that---apparently for numerical reasons related to the relaxation of amplitudes in the LDM iteration---erroneous large peaks occur in the summands of the probe update, we clip all amplitudes of the summands to their 95th percentile value. These peaks also occur in classic DM but are quickly averaged away there due to all exit waves being processed at once and no exit waves being frozen. After these first two corrections, the reconstruction is more stable, see \cref{fig:alg-progression}(c), but the result is still subpar. Thus, we propose to \textbf{3)} pre-reconstruct the central region of the object using the first 20 exit waves in our scan with 200 iterations of DM1. This provides a good initial guess of the probe, the partial object, and the first few exit waves.
When using all three techniques, we already get good reconstructions of the objects, see \cref{fig:alg-progression}(d). We performed the same analysis as in \cref{sec:exp:optimal-params} to determine the best setting for $\iterstotal$ and $\buffersize$, which we found to be $\iterstotal=10$ and $\buffersize=10$ here. Finally, we found that we can get another significant increase in quality by \textbf{4)} using a hybrid algorithm scheme, replacing the 10 LDM1 iterations per scan index with 8 iterations of LDM1 and 2 subsequent iterations of LER. As shown qualitatively in \cref{fig:alg-progression}(e) and quantitatively in \cref{tab:metrics-final}, this final method achieves excellent reconstructions and on average performs significantly better than classic DM1.

\section{Conclusion}
\label{sec:conclusion}
In this work, we have devised and tested novel schemes for the live reconstruction of ptychographic data using a fixed computational budget, by proposing a generic method to extend the established projection-based algorithms to this live setting. Our work enables direct feedback for users during experiments as well as reconstruction of arbitrary-sized objects under limited computational resources and can serve as a basis for adaptive scanning methods similar to \cite{edeAdaptivePartialScanning2021}. We have shown that our proposed methods can achieve reconstructions superior to their non-live counterparts. This holds especially in the setting where the probe has previously been well-characterized, making our methods well-suited to larger-scale experimental settings such as ptycho-tomography. For future work, it should be worthwhile to explore position correction schemes for live reconstruction, adaptive scanning techniques based on our methods, extensions for techniques such as Fourier ptychography and near-field ptychography, and evaluations for experimental data.

\clearpage
\section{References}
\label{sec:refs}
\atColsBreak{\vskip5pt}
\printbibliography[heading=none]

\end{document}